\documentstyle[11pt,moriond_nobox,epsfig]{article}

\def\Journal#1#2#3#4{{#1} {\bf #2}, #3 (#4)}

\def\PLB{{\em Phys. Lett.}  B}

\def\PRD{{\em Phys. Rev.} D}
\def\ZPC{{\em Z. Phys.} C}

\def\EPJC{{\em Eur.Phys.J.} C}

\def\be{\begin{equation}}
\def\ee{\end{equation}}
\def\bea{\begin{eqnarray}}
\def\eea{\end{eqnarray}}

\begin{document}
\vspace*{3.5cm}
\title{Diffractive structure functions from HERA}

\author{Vincenzo Monaco}

\address{Universit\`a di Torino and 
 INFN, Via Giuria 1, 10125 Torino, Italy}

\maketitle

\abstracts{
Recent results on diffraction at HERA
are presented on behalf of the H1 and ZEUS Collaborations, 
focusing on the measurement of the inclusive  
diffractive cross section and on the study of two and three 
jets produced in diffractive $ep$ scattering. 
}

\section{Inclusive diffraction}\label{sec:intro}

At HERA, a significant fraction of neutral-current deep inelastic scattering 
(DIS) events is characterized by a large rapidity gap between the hadronic 
state measured in the detector and the proton direction. 
These events result predominantly from diffractive dissociation 
of the virtual photon;
the lack of QCD radiation in the proton direction is 
interpreted in terms of the exchange of a colour-singlet in the $t$-channel.
In Regge theory~\cite{regge} the colourless system exchanged
at high energy is a 
universal trajectory with the quantum numbers of the vacuum, the Pomeron 
($I\!P$), introduced to describe the energy dependence of the total cross 
sections in hadron-hadron scattering.
In perturbative QCD (pQCD), the vacuum exchange is modelled as a two-gluon 
system that develops into a
gluon ladder between the virtual photon and the proton.
pQCD calculations are possible for 
diffractive processes where a hard scale is present: production of 
high mass quarks or high momentum jets, and, 
at HERA, 
$ep$ scattering
mediated by photons with high virtuality $Q^2$.

In a diffractive DIS event, 
$e(k)p(P)\rightarrow e(k^{\prime})X Y(P^{\prime})$, the photon dissociates into
 the hadronic state X, 
while the proton remains 
intact in the final state or dissociates in a low mass hadronic system Y 
that generally escapes undetected in the beam pipe.
The two hadronic systems
X and Y remain distinct in the final state.

In addition to the usual DIS variables, $Q^2=-q^2=-{(k-k^{\prime})}^2$, 
$y={(P \cdot q)}/{(P \cdot k)}$,  $x=Q^2/{(2P \cdot q)}$,
other quantities are used to describe the kinematics of a diffractive DIS event:
\begin{equation}
x_{I\!P}=\frac{q(P-P^{\prime})}{qp}\simeq \frac{Q^2+M_{X}^2}{Q^2+W^2},
  \hspace{.5cm}
t=(P-P^{\prime})^2, \hspace{.5cm}
\beta=\frac{Q^2}{2q\cdot (P-P^{\prime})}\simeq\frac{Q^2}{Q^2+M_X^2};
\end{equation}

\noindent
where $M_X$ is the invariant mass of the dissociative photon system, $W^2=(q+P)^2$ is the squared $\gamma^*p$ centre of mass energy and $\beta=x/x_{I\!P}$.
In resolved-Pomeron models, where a particle-like Pomeron is assumed to be
emitted from the 
proton and the photon interacts with a partonic component of the Pomeron,
$x_{I\!P}$ is the fraction of the proton momentum carried by the Pomeron, and 
$\beta$ is the fraction of the Pomeron momentum carried by the parton interacting with the virtual photon. 

The QCD factorisation theorem proven for hard diffractive 
scattering~\cite{factor}
 permits the definition of the differential cross section for the reaction 
$ep\rightarrow eXY$ in terms  
of a diffractive structure function $F_2^D$ that depends on universal 
diffractive parton distributions:

\begin{equation}
  \frac{d^4\sigma_{ep}^{diff}}{d\beta dQ^2 dx_{I\!P} dt}=
\frac{2\pi \alpha^2}{\beta Q^4} (1+{(1-y)}^2) F_2^{D(4)}(\beta,Q^2,x_{I\!P},t)
 \label{f2d4}.
\end{equation}
 
\noindent
If $t$ is not measured, the above equation is integrated over $t$ and the 
cross section is defined in terms of the diffractive structure function 
$F_2^{D(3)}(\beta,Q^2,x_{I\!P})$.

The measurements of the inclusive diffractive cross section in 
DIS~\cite{dis_1,dis_2,dis_3,dis_4} are compatible with the resolved-Pomeron 
model of Ingelman and Schlein~\cite{ingelman} (IS) where 
the diffractive 
structure function factorises into a Pomeron flux factor, $f_{I\!P/p}(x_{I\!P},t)$, and a Pomeron structure function, $F_2^{I\!P}(\beta,Q^2)$:

\begin{equation}
  F_2^{D(4)}(\beta,Q^2,x_{I\!P},t)=f_{I\!P/p}(x_{I\!P},t)\cdot F_2^{I\!P}(\beta,Q^2).
\label{factor}
\end{equation}

\noindent
The flux factor, describing the probability of finding a Pomeron in the proton as a function of $x_{I\!P}$ and $t$, is parameterised according to Regge theory: $f_{I\!P/p}(x_{I\!P},t)\sim x_{I\!P}^{1-2\alpha_{I\!P}(t)} $.

\begin{figure}[hb]
\vspace{-0.3cm}
\begin{center}
\epsfig{figure=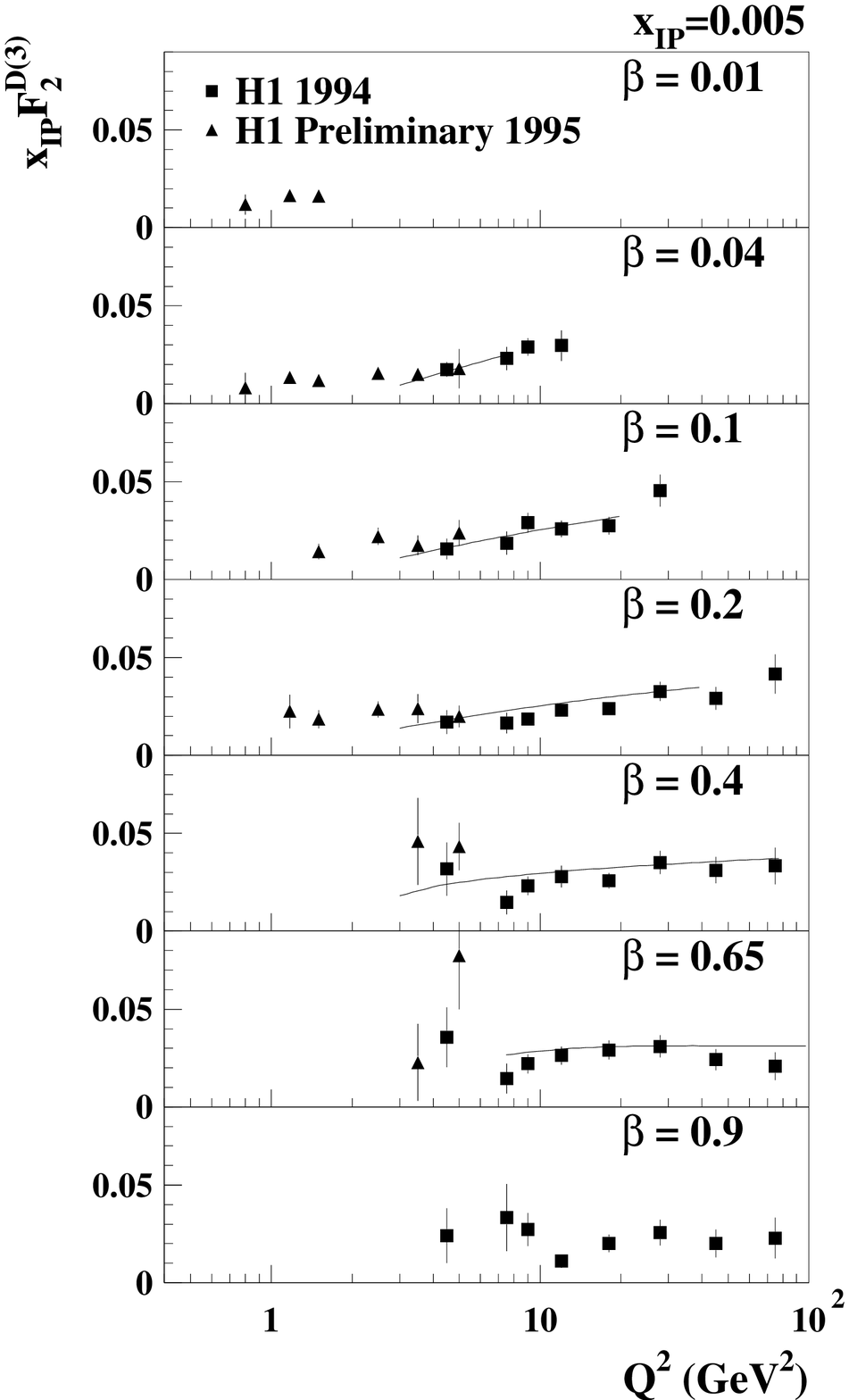,height=3.0in,width=2.8in}
\epsfig{figure=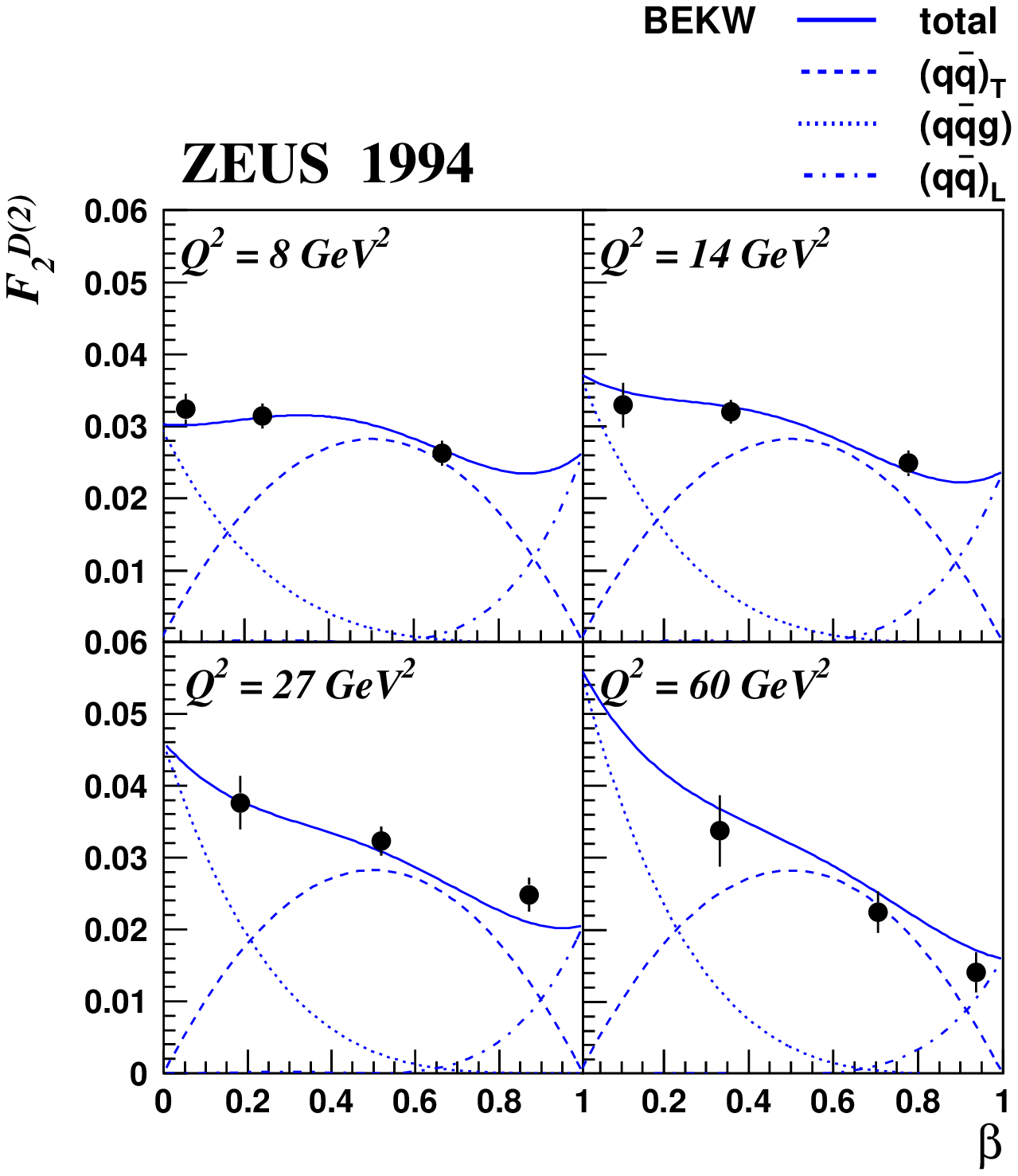,height=3.0in}
\caption{
{\bf Left:}  H1 data on 
${x_{I\!P} F_2^{D(3)}(\beta, Q^2, x_{I\!P})|}_{x_{I\!P}=0.005}=F_2^{D(2)}(\beta,Q^2)$ 
as a function of $Q^2$ for different $\beta$ bins. 
The curve is the result of a DGLAP fit of the parton momentum distributions 
in the pomeron. 
{\bf Right:} Comparison of the ZEUS data on $F_2^{D(2)}$ with a fit based on the BEKW model. } 
\label{fig:f2d}
\end{center}
\end{figure}

In the H1 analysis of inclusive diffractive data~\cite{dis_4} a QCD fit was 
performed in which the parton densities in the Pomeron were evolved according 
to the NLO DGLAP equations. The measured structure 
function $x_{I\!P}F_2^{D(3)}(\beta,Q^2,x_{I\!P})$, shown
in Fig.~\ref{fig:f2d}(left) as a function of $Q^2$ for a fixed $x_{I\!P}$ value and 
for different $\beta$ bins, indicates
a rising scaling violation, persisting also at relatively large values of 
$\beta$. This logarithmic scaling violation is described by the fit, which 
predicts a partonic momentum 
distribution in the Pomeron dominated by gluons.

The approach in which the photon has a pointlike coupling with a partonic 
component of the Pomeron is valid in the proton infinite momentum frame.
Diffractive $\gamma^*p$ scattering can also be studied in the proton rest 
frame, where 
the virtual photon fluctuates into a hadronic state 
(a colour-dipole $q\bar{q}$ or higher Fock states)
 at large 
distances upstream of the proton target.
The $\gamma^*p$ cross section is factorised into the square of an 
effective dipole wave function (calculable in QED) and the cross section for 
the diffractive scattering of the dipole off the proton, which can be calculated in pQCD~\cite{bartels,nikolaev,bialas,bekw}
for high enough values of $Q^2$.

The result of a fit based on a pQCD calculation of the dipole-proton cross 
section (BEKW model~\cite{bekw}) to 
ZEUS data of inclusive diffraction~\cite{zeus_diff} is shown in Fig.~\ref{fig:f2d}(right). 
In the BEKW model the diffractive structure function is the sum of three leading contributions, corresponding to the 
$q\bar{q}$ production from transverse and longitudinal photons, and $q\bar{q}g$ production from transverse photons:
$ F_2^{D(3)}(\beta,Q^2,x_{I\!P})=aF_{q\bar{q}}^T+bF_{q\bar{q}}^L+cF_{q\bar{q}g}^L$.
The fit predicts that the dominant contributions at low $\beta$ comes from the
$q\bar{q}g$ fluctuations of the photon.

The dipole approach has been employed in the {\em ``saturation model''} by Golec-Biernat and W\"usthoff~\cite{sat}. In this model the dipole cross section is 
parameterised as a function of the transverse size of the dipole, 
with the cross section growing from the colour transparency regime at small 
radii (large $Q^2$) to a constant value at large radii.
This model gives a good description of the inclusive proton structure 
function $F_2(x,Q^2)$, including the region of small $Q^2$ values, where a 
partonic model of the Pomeron or pQCD calculations cannot be applied.

\begin{figure}[h]
\begin{center}
\epsfig{figure=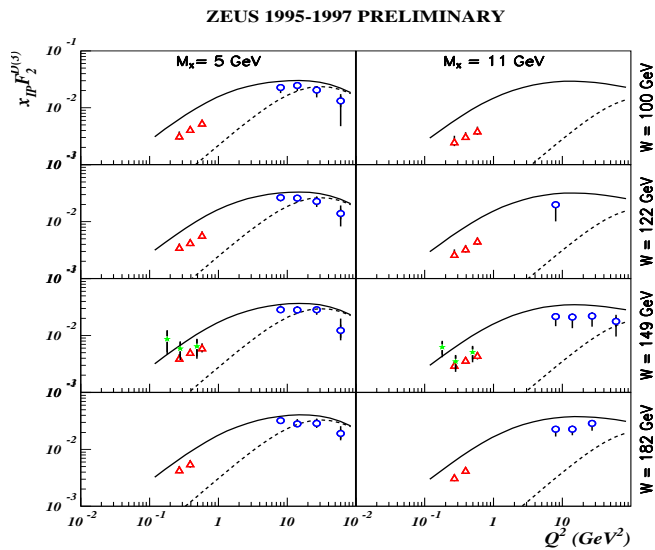,height=3.in,width=3.6in}
\caption{Values of $x_{I\!P}F_2^{D(3)}$ for different $W$ and $M_{X}$ bins as a function of $Q^2$. The stars and the triangles indicate the measurements at low $Q^2$ with the BPC. The higher $Q^2$ points (open cirles) are from previous ZEUS measurements. The curves show the prediction of the ``saturation model'' if only $q\bar{q}$ photon fluctuations are considered ({\em dashed curve}) and with both $q\bar{q}$ abd $q\bar{q}g$ 
fluctuations included ({\em continuous curve}).}
\label{fig:lowq2}
\end{center}
\end{figure}

The diffractive cross section has been recently measured in the transition region between photoproduction and DIS ($0.2<Q^2<0.7$ $\mathrm{GeV^2}$)~\cite{zeus:low_q2}, with the scattered positron detected in the small angle electron calorimeter (BPC) of ZEUS. The $Q^2$ dependence of $x_{I\!P}F_{2}^{D(3)}$ is shown in Fig.~\ref{fig:lowq2} for different $W$ and $M_X$ bins.  A sharp change of the $Q^2$ dependence is observed: while at large $Q^2$ the data do not exhibit a strong $Q^2$ dependence, 
at low $Q^2$ the structure function $x_{I\!P}F_2^{D(3)}$ becomes approximately proportional to $Q^2$.
This behaviour, expected on the basis of the conservation of the electromagnetic current, is described by the saturation model when a contribution of $q\bar{q}g$ fluctuations of the photon is included. The curves shown in 
Fig.~\ref{fig:lowq2} are obtained using the same parameters of the model that successfully describe the inclusive $F_2$ data.

\section{Diffractive jet production}

To further improve the understanding of diffraction, studies of the diffractive final state are performed at HERA.
In particular, final states containing high transverse momentum ($p_T$) jets are calculable in pQCD, and yield direct constraints on the shape of the gluon distribution of the Pomeron.

In the H1 analysis~\cite{h1_2j} DIS events with a large rapidity gap and at least 2 jets with 
$p_{T}> 4$ GeV were found using 
the cone-jet algorithm. 

\begin{figure}[t]
\begin{center}
\epsfig{figure=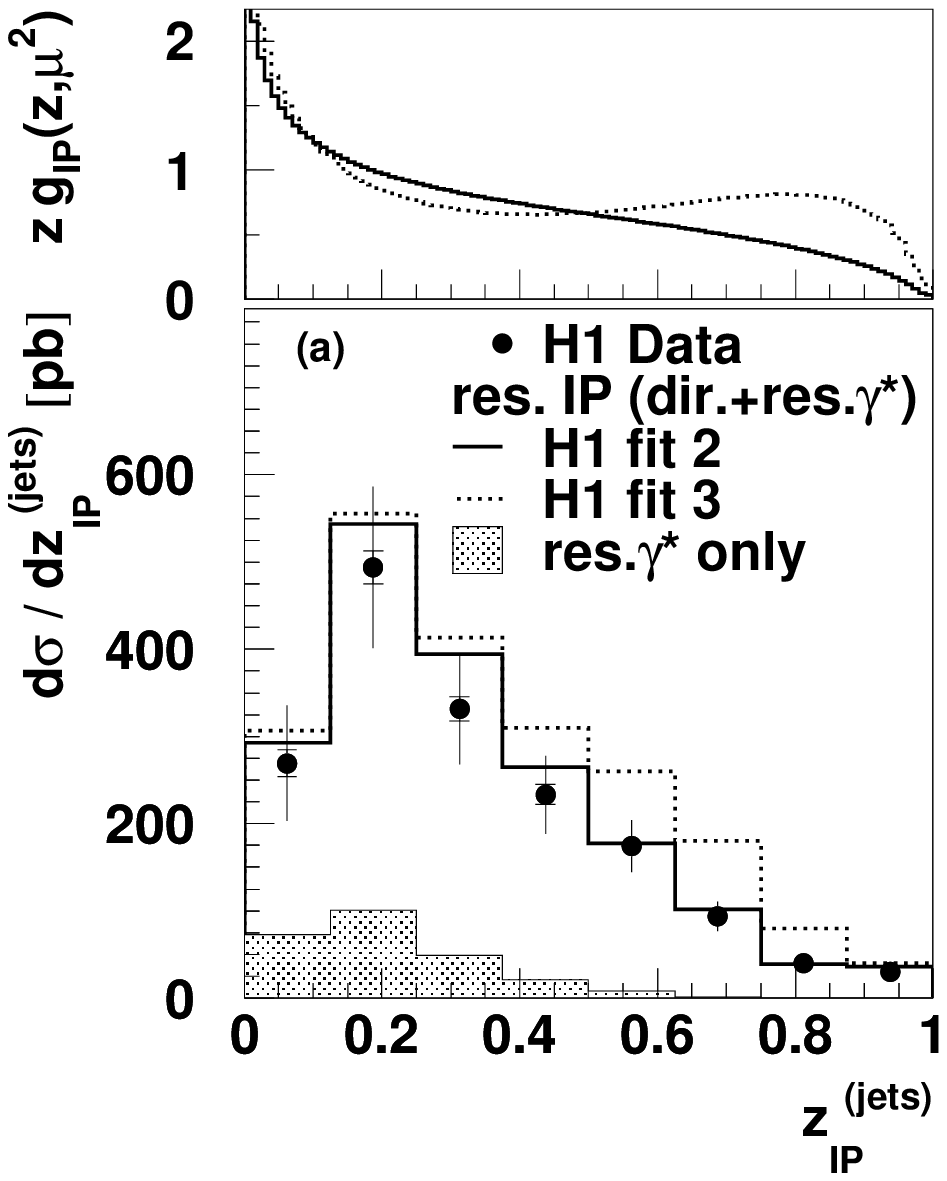,height=3.2in}
\epsfig{figure=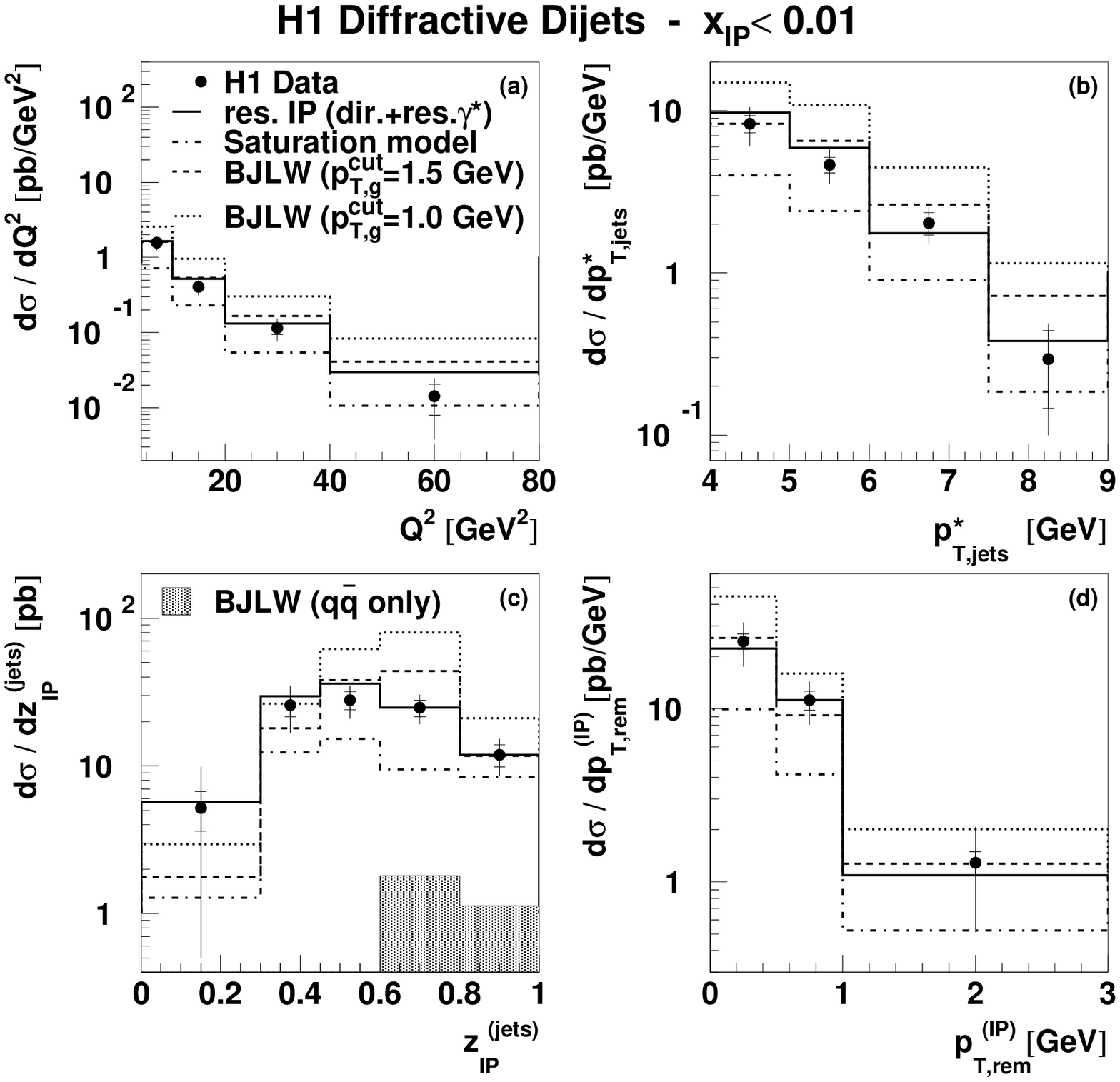,height=3.2in}
\caption{Di-jet production at H1. {\bf Left: } diffractive dijet cross section as a function of $z_{I\!P}^{jets}$, compared with the prediction of the Ingleman-Schlein model for different gluon distributions. {\bf Right: } differential diffractive dijet cross sections in the restricted kinematic range $x_{I\!P}<0.01$ as a function of {\em (a)} $Q^2$, {\em (b)} $p_{T,jets}^*$ ($p_T$ of the jets calculated with respect to the $\gamma^*p$ axis), {\em (c)} $z_{I\!P}^{jets}$
(described in the text), {\em (d)} $p_{T,rem}^{(I\!P)}$ (the summed transverse momentum of the final state particles in the Pomeron hemisphere not belonging to the two jets). }
\label{fig:2j_zp}
\end{center}
\end{figure}

In Fig.~\ref{fig:2j_zp}(left) the quantity 
$z_{I\!P}^{jets}=(Q^2+M_{12}^2)/(Q^2+M_X^2)$, representing the fraction of the hadronic energy in the final state contained in the two jets, is compared 
with the prediction 
from the IS model based on different sets of Pomeron gluon 
distributions obtained from the leading order DGLAP analysis of $F_2^{D(3)}$ 
by H1. The data favour a Pomeron dominated by gluons with a gluon momentum 
distribution that is relatively flat in $z_{I\!P}^{jets}$.

Several differential distributions are compared in Fig.~\ref{fig:2j_zp}(right) with the IS model (res.IP in the figure), the saturation model and a pQCD calculation based on a dipole approach (BJLW model~\cite{bartels}). 
In the IS model the same flux factor and Pomeron 
structure functions describe both the inclusive diffractive data and the 
dijet distributions, consistent with the factorization hypothesis expressed by Eq.~\ref{factor} and with the universality of the Pomeron parton distributions in $ep$ scattering.

For diffractively scattered $q\bar{q}$ photon fluctuations, a distribution 
peaked at $z_{I\!P}\sim 1$ is expected in Fig.~\ref{fig:2j_zp}({left}); 
the low values 
of $z_{I\!P}$ measured in the two-jet sample imply the dominance of 
$q\bar{q}g$ over $q\bar{q}$ scattering in the proton rest frame picture.
Futhermore, the high gluon content of the Pomeron favours a picture in 
which, in the proton infinite momentum frame, the dominant contribution is a 
Boson-Gluon Fusion process, with two jets and a gluon remnant in the final 
state. This picture is consistent with the $F_2^D$ data shown in Fig.~\ref{fig:f2d}(right) where at low 
$\beta$ (or high $M_{X}$) the contribution of $q\bar{q}g$ photon fluctuations 
is expected to dominate. In these events the gluon is emitted almost 
collinear to the $\gamma^* I\!P$ axis in the Pomeron direction, with small 
$p_T$, while the two quarks are emitted in the photon emisphere. 

In order to confirm the validity of this picture, three-jet production in 
diffractive DIS was studied by both H1~\cite{h1_2j} and ZEUS~\cite{zeus_3j} collaborations. 
In the ZEUS analysis the exclusive $k_T$-algorithm~\cite{kt_alg}
 was used in the centre 
of mass of the observed hadronic final state to select jet configurations 
aligned with respect to the $\gamma^*I\!P$ axis.
The measurement was restricted 
to high-mass final states ($M_X> 23$ GeV), in order to separate clear three-jets topologies, and diffractive events were selected by requiring a large rapidity gap in the outgoing proton direction.

\begin{figure}[h]
\begin{center}
\vspace{-1cm}
%\rule{5cm}{0.2mm}\hfill\rule{5cm}{0.2mm}
%\vskip 2.5cm
%\rule{5cm}{0.2mm}\hfill\rule{5cm}{0.2mm}
\epsfig{figure=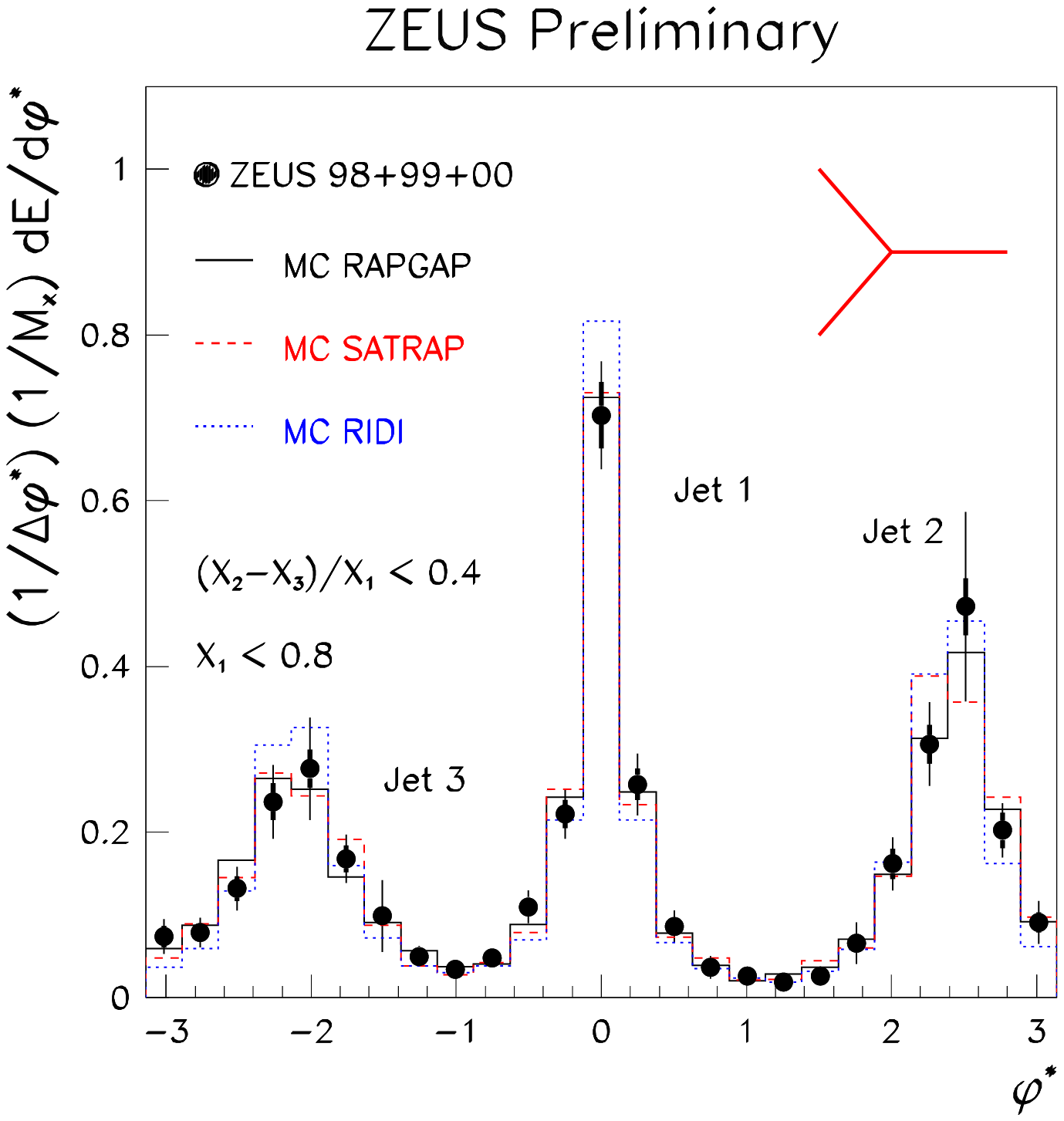,height=3.2in}
\hspace{-1.5cm}
\epsfig{figure=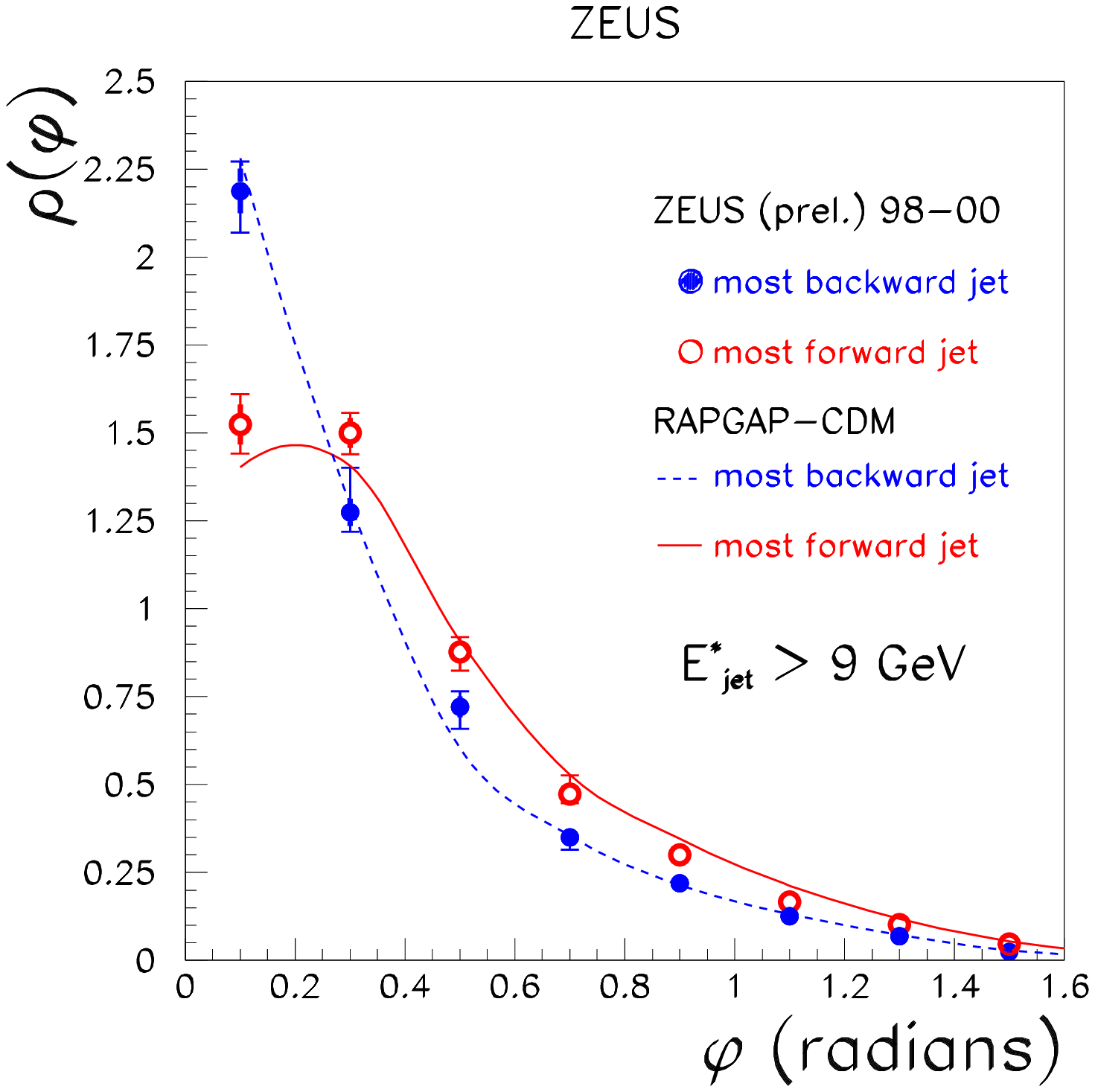,height=3.2in}
\vspace{-1cm}
\caption{{\bf Left: } Energy distribution normalized to the total invariant mass $M_{X}$ as a function of the azimuthal angle $\varphi^*$ defined in the plane of the two most energetic jets. {\bf Right:} Differential jet shape for the most forward and most backward jet in three jets events with $E^{jet}>9$ GeV.}
\label{fig:en_fl}
\end{center}
\end{figure}

The energy flow, measured with respect to the azimuthal angle $\varphi^*$ 
in the event plane defined by the two most energetic jets, is shown in 
Fig.~\ref{fig:en_fl}(left). A clear three-jet structure, reproduced by different MC models, is observed.

The shapes of the jets were studied with the aim of distinguishing between 
quark and gluon-initiated jets. The differential jet shape $\rho(\varphi)$, defined  as the 
average fraction of the energy of the jet which lies in an annulus of inner 
angular distance $\varphi-\delta\varphi/2$ and outer angular distance 
$\varphi+\delta\varphi/2$ around the jet axis, is shown in 
Fig.~\ref{fig:en_fl}(right) as a function of $\varphi$ for the most-forward 
and most-backward jet, where the 
forward direction is defined by the Pomeron direction. The measured jet in 
the Pomeron direction is broader than the jet in the photon direction, as 
expected for a jet initiated by a gluon preferentially emitted in the Pomeron direction. The jet shape distribution is reproduced 
by the RAPGAP MC model~\cite{rapgap}, based on the resolved-Pomeron model 
with a Pomeron dominated by gluons.

\section{Summary}

Recent results on diffraction from HERA have been reviewed.
The cross section has been presented in terms of the diffractive structure 
function $F_2^D$ and the jet-structure of the hadronic final state $X$ has been discussed. The data are described by models based on the Ingelman and 
Schlein approach, where the parton distributions in the Pomeron are probed, in the proton infinite momentum frame, by a pointlike virtual photon.  
 
The data are also described in a consistent way by
pQCD calculations in the proton rest frame, where the photon fluctuates 
into a colour dipole that interacts with the proton via the exchange of a gluon ladder; 
a dipole approach with the dipole-proton cross section parameterised 
as a function of the dipole radius describes the transition to low $Q^2$.

In both points of view, three partons in the final state are expected 
at low $\beta$ (or, equivalently, in events with high hadronic mass $M_X$), 
with a low-$p_T$ gluon emitted preferentially in the Pomeron direction. 
Three jets events were observed at HERA, with a topology consistent with this 
picture.


\newpage

\begin{thebibliography}{99}

\bibitem{regge} P.D.B.Collins in {\em ``An introduction to Regge theory and high energy physics''}, Cambridge Univ.Press, Cambridge (1977).

\bibitem{factor} J.C.Collins, \Journal{\PRD}{57}{3051}{1998}.

\bibitem{dis_1} H1 Collab., T.Ahmed et al., \Journal{\PLB}{348}{681}{1995}.

\bibitem{dis_2} ZEUS Collab., M.Derrick et al., \Journal{\ZPC}{68}{569}{1995}.

\bibitem{dis_3} ZEUS Collab., J.Breitweg et al., \Journal{\EPJC}{1}{81}{1998}.

\bibitem{dis_4} H1 Collab., C.Adloff et al., \Journal{\ZPC}{76}{613}{1997}.


\bibitem{ingelman} G.Ingelman and P.E.Schlein, \Journal{\PLB}{152}{256}{1985}.

\bibitem{bartels} J.Bartels, H.Jung and M.W\"usthoff, {\em Eur.Phys.J} {\bf 11}, {111} {(1999)}.

\bibitem{nikolaev} M.Genovese and N.N.Nikolaev, {\em J.Exp.Theor.Phys.} {\bf 81}, 633 (1995).


\bibitem{bialas} A.Bialas, R.Peschanski and C.Royon, \Journal{\PRD}{57}{6899}{1998}.

\bibitem{bekw} 
J.Bartels, C.Ewerz, H.Lotter and M.W\"usthoff, \Journal{\PLB}{386}{389}{1996}; \\
H.Lotter, \Journal{\PLB}{406}{171}{1997}; \\
J.Bartels, J.Ellis, H.Kowalski and M.W\"usthoff, \Journal{\EPJC}{7}{443}{1999}.

\bibitem{zeus_diff} ZEUS Coll., J.Breitweg et al., \Journal{\EPJC}{6}{43}{1999}.
                    

\bibitem{sat} K.Golec-Biernat and M.W\"usthoff, \Journal{\PRD}{59}{014017}{1999}; \\
             K.Golec-Biernat and M.W\"usthoff, \Journal{\PRD}{60}{114023}{1999}.

\bibitem{zeus:low_q2} ZEUS Coll., J.Breitweg et al., paper 435 submitted to the XXXth International Conference on High Energy Physics, Osaka, Japan (August 2000).

\bibitem{h1_2j} H1 Collab., C.Adloff et al., DESY 00-174, hep-ex/0012051.

\bibitem{zeus_3j} ZEUS Coll., J.Breitweg et al., paper 433 submitted to the XXXth International Conference of High Energy Physics, Osaka, Japan (August 2000).

\bibitem{kt_alg} S.Catani et al., \Journal{\PLB}{269}{432}{1991}

\bibitem{rapgap} H.Jung, {\em Comp.Phys.Comm.} {\bf 86} 147 (1995).

\end{thebibliography}
\end{document}